\documentclass{emulateapj}
\usepackage{graphicx}
\usepackage{times}
\usepackage{mathptm}

\def\spose#1{\hbox to 0pt{#1\hss}}
\def\approxgt{\mathrel{\spose{\lower 3pt\hbox{$\sim$}}
        \raise 2.0pt\hbox{$>$}}}

\slugcomment{}

\shorttitle{Substructure depletion by the disk }
\shortauthors{D'Onghia et al.}

\begin{document}


\title{Substructure depletion in the Milky Way halo by the disk}


\author{Elena D'Onghia\altaffilmark{1,2,3}, Volker Springel\altaffilmark{4},
Lars Hernquist\altaffilmark{1},
 Dusan Keres\altaffilmark{1}}
\affil{1-Harvard-Smithsonian Center for Astrophysics,
60 Garden Street, Cambridge, MA 02138 USA\\
4-Max-Planck-Institute for Astrophysics,
Karl-Schwarzschild-Str. 1, 85740 Garching, Germany
}




\altaffiltext{2}{Marie Curie fellow; edonghia@cfa.harvard.edu}
\altaffiltext{3}{Institute for Theoretical Physics, University of Zurich, 
Winterthurerstrasse 190
CH-8057 Zuerich, Switzerland}


\begin{abstract}

We employ numerical simulations and simple analytical estimates to argue
that dark matter substructures orbiting in the inner regions of the
Galaxy can be efficiently destroyed by disk shocking, a dynamical
process known to affect globular star clusters.  We carry out a set of
fiducial high-resolution collisionless simulations in which we
adiabatically grow a disk, allowing us to examine the impact of the disk
on the substructure abundance.  We also track the orbits of dark matter
satellites in the high-resolution Aquarius simulations and analytically
estimate the cumulative halo and disk shocking effect.  Our calculations
indicate that the presence of a disk with only 10\% of the total Milky
Way mass can significantly alter the mass function of substructures in
the inner parts of halos.  This has important implications especially
for the relatively small number of satellites seen within $\sim$30 kpc
of the Milky Way center, where disk shocking is expected to reduce the
substructure abundance by a factor of ~2 at 10$^9$ M$_{\odot}$ and
  ~3 at 10$^7$ M$_{\odot}$.  The most massive subhalos with 10$^{10}$
  M$_{\odot}$ survive even in the presence of the disk.  This suggests
that there is no inner missing satellite problem, and calls into
question whether these substructures can produce transient features in
disks, like multi-armed spiral patterns.
Also, the depletion of dark matter substructures through shocking on the
baryonic structures of the disk and central bulge may aggravate
the problem to fully account for the observed flux anomalies in
gravitational lens systems, and significantly reduces the dark matter
  annihilation signal expected from nearby substructures in the inner
  halo.

\end{abstract}


\section{Introduction}

In the cold dark matter (CDM) scenario, structure grows hierarchically, with
small objects collapsing first and continuously merging to form larger and
larger bodies \citep{WR78}.  Numerical simulations have tracked the evolution
of this process, beginning at early times with objects having masses
comparable to that of the Earth \citep{D05}, and progressing through many
orders of magnitude in mass to the scales of galaxies, galaxy clusters, and
cosmic large-scale structure \citep[e.g.][]{Ghigna2000,Spr2005}.  In the Milky
Way, the cumulative number $N$ of dark matter satellites of a given mass
$M_{\rm{sat}}$ is predicted to scale with the subhalo mass as $N\propto
M^{-0.9}_{\rm{sat}}$ \citep{Setal08}.  Because the subhalos hosting the
satellite galaxies formed early, when the Universe was dense, the smaller
structures are thought to be resilient to tidal disruption and the simulations
predict that $\sim 100$ with maximum circular velocity larger than $20\, {\rm
  km \,s^{-1}}$ should survive to the present day for a halo like that of the
Milky Way \citep{K99,M99,Detal08,Setal08,K09}.

However, these estimates of substructure abundance at different
locations within a halo, from the solar neighborhood to the outskirts
of the Milky Way, are based on simulations that include only dark
matter and do not account for ordinary baryonic material \citep[but
see][]{Dolag2008}.  This means that any dynamical coupling between
dark matter substructures and the luminous components of galaxies has
been ignored, even though this may alter the substructure abundance.
A precise understanding of the impact of baryons on the substructure
abundance is important to correctly assess the relevance of
substructures for the evolution of galaxies, and for attempts to
detect dark matter through gravitational lensing or annihilation
radiation.

For example, recent studies suggest that cosmological infall of 
dark satellites through the disk of a typical galaxy could induce bars
and multi-armed spiral patterns in disks \citep{Dub08}.  According to
these simulations, the main agent producing transient features would
be satellite passages through the inner part of the disk.  Indeed, it
is speculated that the passage of a satellite through the disk induces
a localized disturbance that could grow by the mechanism termed 
swing amplification \citep{T81}.  Because the tidal effects of the
satellites are generally small, this process is distinct from
interactions thought to be responsible for grand-design spirals like
M51 \citep{Kaz09}.  The mechanism proposed by \citet{Dub08} is closer
in spirit to the original ideas suggested by 
\citet{JT66} in which an object within a disk excites a response that
is subsequently amplified. Unlike the interactions involving
from large visible satellite galaxies,   we are more interested  in the 
process produced by the continuing
bombardment of a galactic disk by the dark matter halo inhomogeneities that will
keep on exciting ragged spiral structures. 

Here, we investigate the coupling between dark matter substructures
and the luminous components of galaxies.  We consider tidal processes
that lead to a depletion of the substructures orbiting in the inner
regions of a galaxy by the presence of a disk and we discuss its
astrophysical implications.

Naively, one might expect that the luminous components of galaxies
should play only a minor role in affecting substructures because most
of the mass is in the dark matter.  However, some gravitational
interactions between bodies, especially those involving tidal forces,
depend not on the relative masses of objects, but on their relative
densities.  Whereas the dark matter is thought to be nearly
collisionless, baryons can dissipate energy and angular momentum and
collapse to high densities.  This must have been the case for the
baryons in the disk of the Milky Way, which is far more concentrated
in its inner parts than the dark matter that makes up the surrounding
smooth halo.  Thus, it is plausible that the baryons could have an
impact on the dark matter substructures despite their small
contribution to the overall matter density of the Universe, motivating
our investigation.

Specifically, we carry out a number of fiducial numerical experiments
in which we adiabatically grow a disk in a high-resolution
collisionless simulation of a Milky-Way sized halo. This enables us
to examine the impact of e.g.~disk shocking on dark matter
substructures that pass through the disk.  We also follow the orbits
of substructures in the high-resolution dark matter simulations of the
Aquarius project, allowing us to identify the parameters and frequency
of encounters with an assumed disk in the halo.  This provides an
independent check of our isolated disk-growth simulations, and also
enables an investigation of the consequences of different modeling
assumptions.

In \S\ref{intro}, we discuss the numerical set-up of our disk-growth
simulations, and in \S\ref{results} we present the results.  In
\S\ref{aquarius} we compare with analytic estimates of the disk
shocking effect based on tracking satellite orbits in cosmological
simulation of halo formation.  Finally, we discuss the implications of
our results in \S\ref{discussion}.

\begin{figure*}
\epsscale{1.0}
\plotone{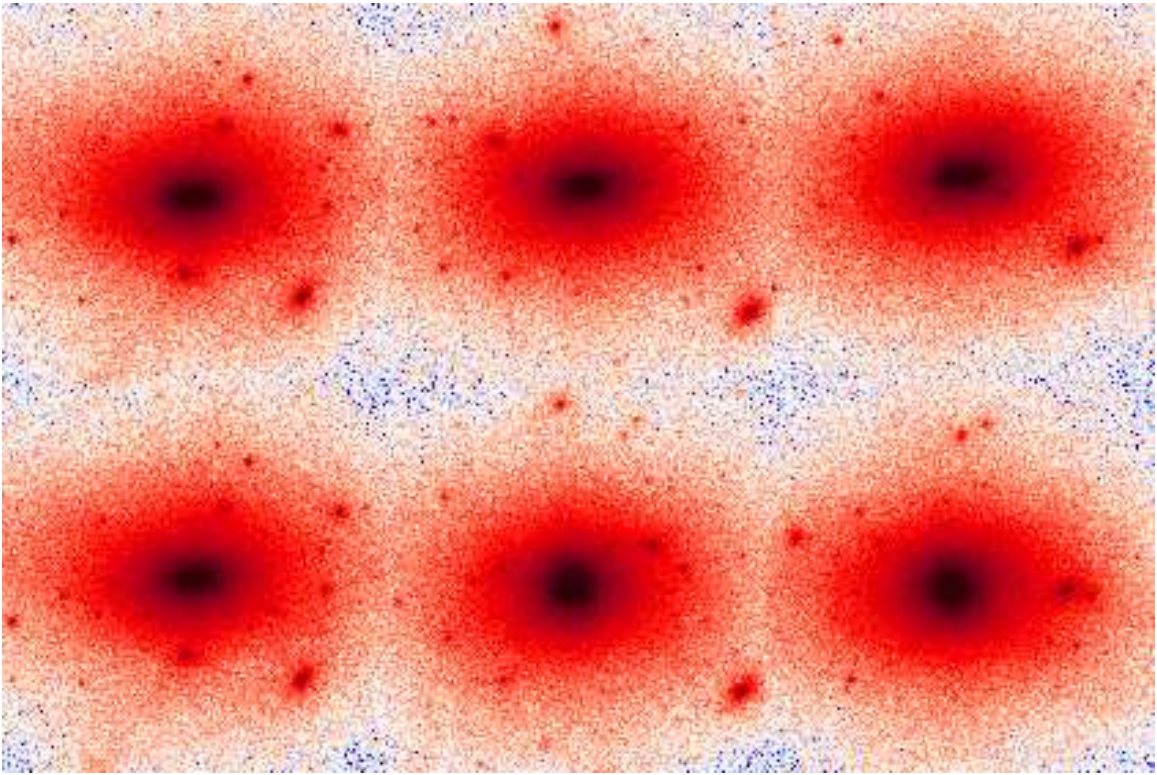}
\caption{Dark matter distribution within 60 kpc of the center of
  the dark matter halo in two simulations,
  each at three different times.  The columns are the initial state
  (left column; identical in both cases), after 3.7 Gyrs (middle
  column), and after 5.5 Gyrs (right column).  Top row is a simulation
  with no disk, bottom row is one where a disk with mass 0.1 times the
  dark halo was included.  For the experiment with a disk, the disk was
  grown adiabatically over the initial 2 Gyrs to its final
  mass. \label{FigHalos}}
\end{figure*}

\section{NUMERICAL METHODS}   \label{intro}

Our simulations were performed with {\small GADGET3}, a parallel
TreePM code developed to compute the evolution of dissipationless dark
matter systems.  A detailed description of the code is available in
the literature \citep{S05,Setal208}.  Here, we note its essential
features. 

{\small GADGET3} is a cosmological code in which the gravitational
field on large scales is calculated with a particle-mesh (PM)
algorithm, while the short-range forces are provided by a tree-based
hierarchical multipole expansion, such that an accurate and fast
gravitational solver results.  The scheme combines the high spatial
resolution and relative insensitivity to clustering of tree algorithms
\citep{BH88} with the unmatched speed and accuracy of the PM method to
calculate the long range gravitational field.  Pairwise particle
interactions are softened with a spline of scalelength $h_s$, so that
they are strictly Newtonian for particles separated by more than $h_s$
\citep{HK89}.  For our applications, the gravitational softening
length is kept fixed in comoving coordinates throughout the evolution
of our halos.

\subsection{Setting the initial conditions}

Presently, galaxy formation is not well-understood and so it is not
possible to perform an {\it ab initio} cosmological simulation that
combines the evolution of dark matter and baryons and produces an
object that closely resembles the Milky Way.  For this reason, we
instead employ simulations in which we follow the growth and
subsequent gravitational collapse and virialization of dark matter
halos self-consistently, but include the gravitational contribution
from luminous galaxies in an idealized manner.  In particular, we
represent the baryons of a galaxy as a fixed disk potential, added to
a dark matter simulation with parameters and temporal behavior chosen
to match observational and theoretical constraints.  This enables us
to study the dynamical coupling between luminous galaxies and
substructures in their dark matter halos.

In our procedure, we first evolve cosmological simulations with dark
matter only and locate halos with properties at the present day
similar to those of the Milky Way.  The halo described in detail below
was originally selected within a low resolution, dark matter only
simulation run in a concordance, $\Lambda$-dominated universe with
parameters $\Omega_m = 0.27$, $\Lambda=0.73$, $h = 0.7$, $\sigma_8 =
0.8$, and $\Omega_b=0.044$.  These cosmological parameters are
consistent with the current set of constraints within their
uncertainties, in particular those from the WMAP 1- and 5-year data
analysis \citep{Komatsu09}.

The large dynamic range involved in cosmological simulations aimed at
resolving the scales of halos like the one presented here calls for
techniques that concentrate the computational power on the object of
interest.  This is achieved by a zoom-technique, where a large-scale
simulation is done at low resolution and the relevant regions are
identified within it, e.g.~a Milky Way-sized halo such as in this
work.  The simulation can then be repeated with much higher resolution
in this region.

The size of the box we employed, $10\, h^{-1}{\rm Mpc}$ on a side, is
quite small and starts to become non-linear at the present epoch,
but it is still sufficient to provide realistic torques and 
accurate formation histories of Milky-Way sized halos.
The virial mass of the halo we selected at $z = 0$ to be re-simulated at
higher resolution is $5.5 \times 10^{11}\,h^{-1}$M$_{\odot}$ (the virial mass
is here measured within a radius $R_{\rm vir} \sim 160\,h^{-1}{\rm kpc}$,
enclosing an overdensity of $\sim 96$ times the critical density \citep{BN98}).

Dark matter particle masses in the high resolution regions are
$5.5\times 10^5\,h^{-1}M_{\odot}$, and the force resolution, i.e.~the
gravitational softening, is $200\,h^{-1}{\rm pc}$. In total there are
$\sim 10^6$ particles within the virial radius.  With our choices for
particle number and softening, the smallest subhalos resolved have
typical masses of $\sim 10^7\,M_{\odot}$.  The chosen halo has a
merger history and spin parameter reasonably representative of the
global population of halos in equilibrium \citep{DN07}.  The halo was
selected with the only criterion that the redshift of the last major
merger was $z\approx 1.5$ and that there are no halos of similar or
larger mass within a few virial radii (a major merger is defined here
as having a 4:1 mass ratio). The halo spin parameter is $0.04$ at $z =
0$, close to the average value $\approx 0.035$ for halos in large
cosmological simulations \citep{Bett07}.

\subsection{Putting a disk inside the galactic halo}

We began the cosmological simulation at high resolution in the region of
interest at an early time and evolved it forward to the present day (and
beyond), saving the state of the system at many intermediate times.  We
extracted the galaxy halo at the present time from the cosmological
simulation, including all particles within 500 $h^{-1}$ kpc of the galactic
center and then repeated the simulations, but adding the potential from a disk
to the halo starting at a specific time.  We repeated this procedure by
varying both the manner in which the galaxy was included as well as its
properties.

We have modeled the disk potential according to \citep{Kuz56}:
\begin{equation}
\phi(R,z)=-\frac{G M}{\sqrt{R^2+(a+\sqrt{z^2+b^2})^2}},
\end{equation}
where $M$ is the disk mass, $a$ is the scale-length of the disk and is
taken to be $4.55 h^{-1}\, {\rm kpc}$, and $b$ is the disk height, assumed to
be $260\,{\rm pc}$. With these choices, this potential matches that of
the Galactic disk \citep{JSH95}.  We have tested different
prescriptions to place the disk inside the halo: we either centered on
the halo center of mass, or on the particle where the halo potential
is a minimum.  In the former method we adopted an iterative method to
locate the maximum density and centered the disk there. The different
procedures gave similar results; for the final runs analyzed in
this paper we adopted the minimum potential approach.

In the simulations with disks, we grew the galaxy slowly
(adiabatically) for $2\,{\rm Gyrs}$ to its full mass to avoid initial
transients.  In our fiducial case, the disk grows to a mass of 1/10th
of that of the halo within its virial radius, but we also performed
different simulations altering the properties of the luminous galaxy.
In particular, we varied the mass of the disk relative to that of the
dark matter halo, its radial extent, and the growth timescale of the
disk.  In general, we see the heating processes becoming stronger as
the mass of the disk or its scale length are increased, consistent
with analytic expectations (see below).

We investigate the abundance of dark matter substructures inside the
halo as measured by the {\small SUBFIND} algorithm
\citep{Springel2001}.  All our substructures consist of particle
groups that are gravitationally self-bound and are overdense with
respect to the local background. Every simulation particle can be part
of only one subhalo. We count substructures down to a minimum of 20
bound particles.

\section{RESULTS FOR DISK-GROWTH SIMULATIONS}  \label{results}

The generic outcome of our experiments is illustrated in
Figure~\ref{FigHalos}, which shows the appearance of halos at various
times for cases in which we ignored the contribution from a disk (top
row) and included it (bottom row).  From left to right, we show the
dark matter distribution initially and then after 3.7 and $5.5\,{\rm
Gyrs}$.  The simulations shown in this figure assumed a disk with a
final mass of $5.5\times 10^{10}\, h^{-1}$ M$_{\odot}$, i.e.~$10\%$ of
the dark mass within the virial radius, appropriate for the Milky Way.

\begin{figure}
\epsscale{1.0}
\plotone{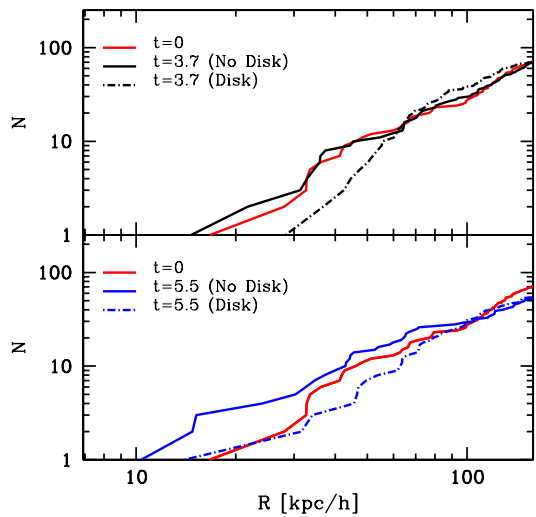}
\caption{Cumulative number of subhalos as a function of distance from
  the center of the host halo, with and without the disk.
 \label{FigRadialAbundnace} }
\end{figure}

\begin{figure}
\epsscale{1.0}
\plotone{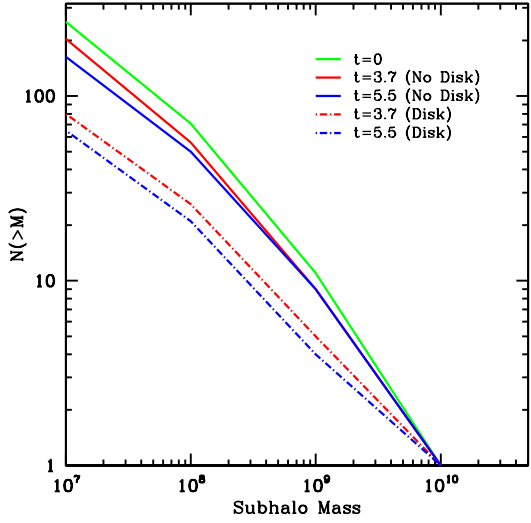}
\caption{Cumulative number of subhalos above a given mass limit,
for the evolved simulations with and without a disk within two virial radii.
At late times, the disk can reduce the substructure 
abundance by nearly a factor of two.
\label{FigCumulativeAbundnace} }
\end{figure}

\begin{figure}
\epsscale{0.5}
\plotone{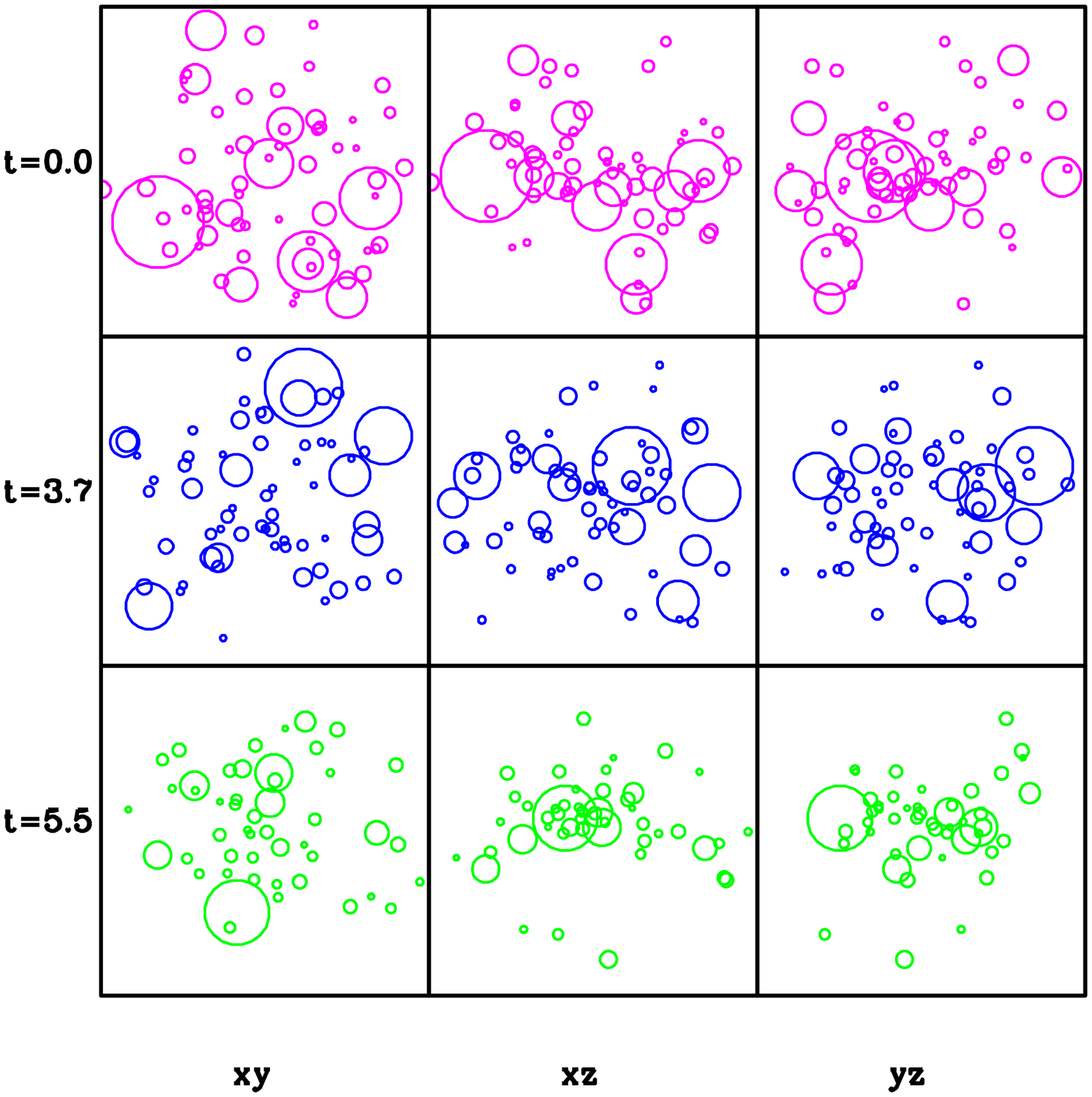}
\epsscale{0.5}
\plotone{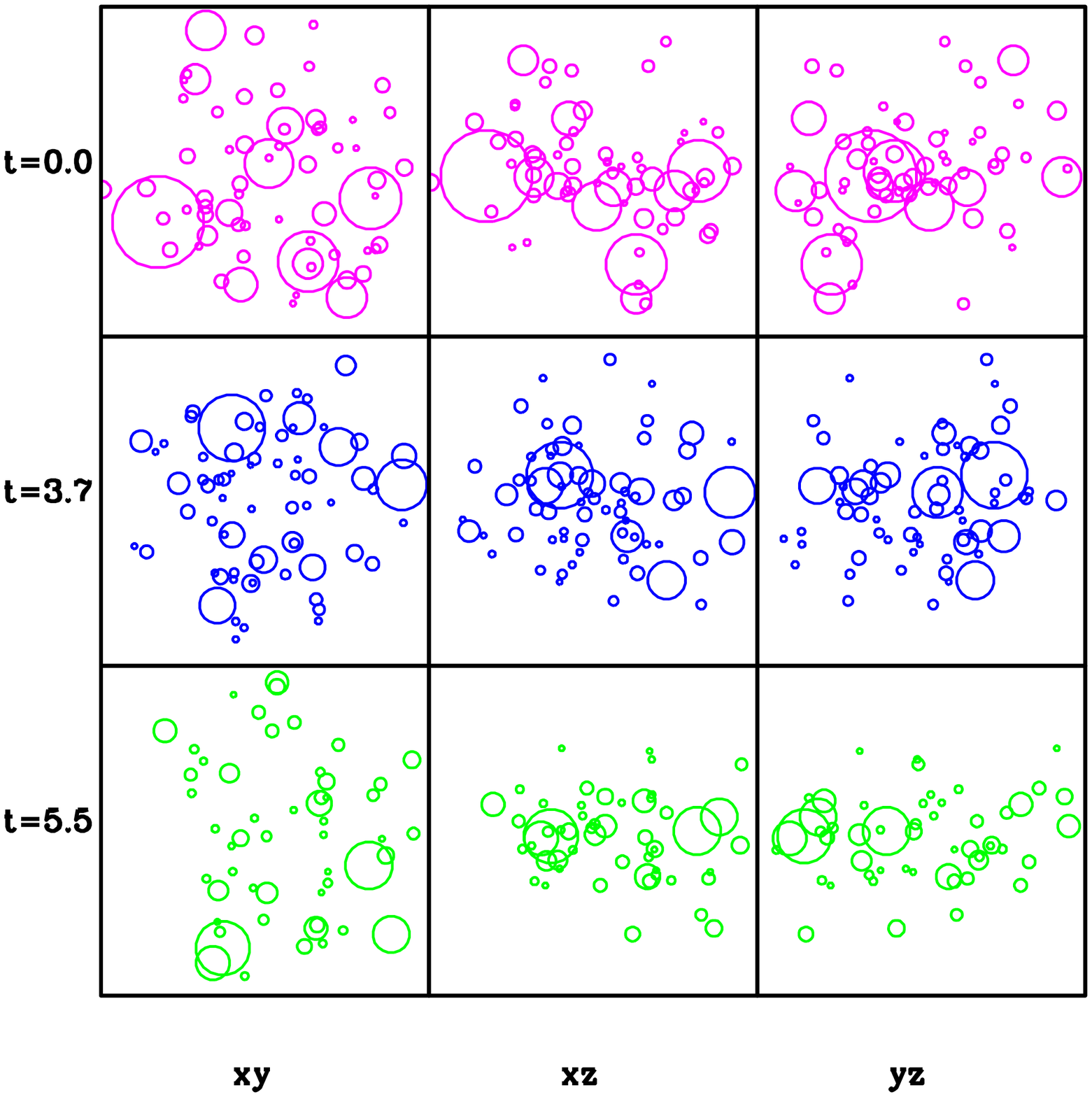}
\caption{Spatial distribution of substructures within $150 h^{-1}{\rm
    kpc}$ of the galaxy center, comparing the case when a disk is not
  included (top set of panels) with the case where disk is included
  (bottom set of panels).  For both, we show the distribution at
  the initial time $t=0$ (top rows) and after $3.7\,{\rm Gyrs}$ and
  $5.5\,{\rm Gyrs}$, respectively (middle and bottom rows).  
  Circle areas are proportional to the substructure
  masses. \label{FigSpatial} }
\end{figure}

\begin{figure}
\epsscale{1.0}
\plotone{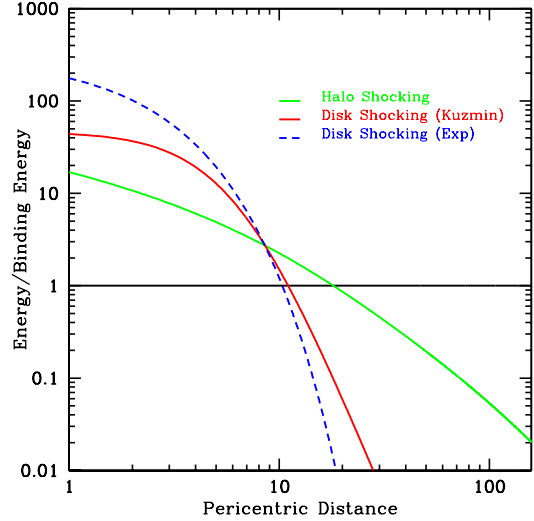}
\caption{Energy of a satellite of $10^7\,$ M$_{\odot}$ as compared to its
  binding energy being affected by disk shocking produced by an exponential
  disk (blue dashed line), by a Kuzmin disk as in our simulations (solid red
  line), and by tidal shocking from the halo (green solid line) for different
  pericentric distances of the satellite orbit. \label{FigEfficiency}}
\end{figure}

As is visually apparent in Figure~\ref{FigHalos}, the dark matter
distribution is significantly influenced by a disk.  Within $2\,{\rm
Gyrs}$ after reaching its final mass, the disk is able to erase most
of the substructures inside the inner few tens of kpc of the halo.
Because this timescale is short compared to the age of the Universe
and the evolutionary timescales of real galactic disks, we expect that
few, if any, low mass substructures would actually be orbiting the innermost
regions of a large spiral galaxy.

This is made more explicit in Figure~\ref{FigRadialAbundnace}, where
we compare the radial cumulative abundance of substructures between
the runs with and without disks.  As time goes by, the depression of
the subhalo function in the disk case increases, in proportion to the
larger number of disk passages that have occurred.  This substructure
depletion is particularly strong in the inner parts of the halo, even
though an effect is also noticeable in the outer parts.  In
Figure~\ref{FigCumulativeAbundnace}, we plot instead the cumulative
substructure count as a function of mass, comparing again the
simulations with and without a disk at two different times after the
start of the simulations. There is up to a factor of $\sim 2$
reduction in the substructure abundance, and the effect is
approximately independent of substructure mass if we discard the
measurement for the most massive bin, which is populated by only one
object and therefore allows no statistical conclusions.

The depletion of the substructure abundance is also reflected in the
spatial distribution of substructures, which we show in
Figure~\ref{FigSpatial} at different times for the cases without a
disk (top) and when a disk with a final mass equal to 10\% of the
final halo mass included (bottom).  The areas of the symbols are
proportional to satellite mass.

Figures~\ref{FigRadialAbundnace}, \ref{FigCumulativeAbundnace} and
\ref{FigSpatial} confirm that a disk accelerates mass loss by
satellites, altering the subhalo mass function.  Overall, the effect
of the disk is to reduce the number of substructures and the masses of
those that survive already after 3.7 Gyrs.  Moreover, with time the
disk also compresses the dark matter distribution in the center,
further contributing to subhalo heating and accelerating the mass loss
of these systems.  A measurement of the halo radial dark matter
density profile showed that the halo contracts and becomes denser by a
factor of 2 within the inner regions owing to the gravitational
potential of the disk.  We also note that the presence of a disk makes
the inner parts of the dark matter halo rounder, an effect that we
further quantify below.

\begin{figure}
\epsscale{1.0}
\plotone{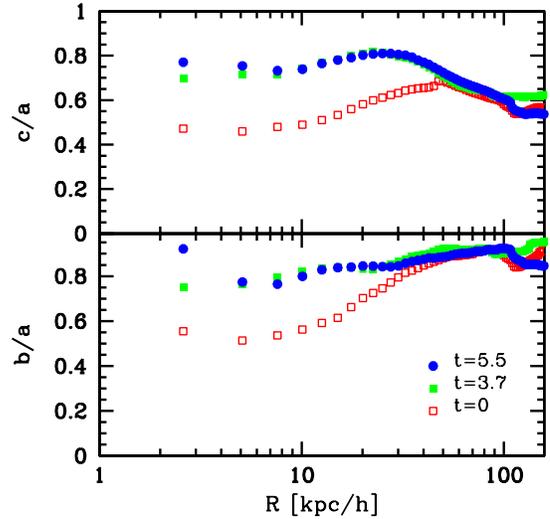}
\caption{Radial dependence of the intermediate to major axis ($b/a$) (bottom
  panel) and minor to major axis ($c/a$) (top panel) ratios of ellipsoids
  fitted to the halo when a disk is included. \label{FigShape}}
\end{figure}

We can identify three physical mechanisms which cause substructures to
lose mass and to eventually be disrupted.  First, as these objects
orbit within the halo-galaxy system, they will be {\it tidally
stripped}. The contribution of the halo to this process has been
included in previous studies \citep{Hay03,Ste04,Read06,Pen08,Mun08} but generally
not that from the luminous galaxy \citep[but see][]{TB01}.  Moreover,
the adiabatic contraction of the halo owing to the disk enhances this
process, an effect that is not accounted for in dark matter only
simulations.

The impact of this effect can be estimated crudely by assuming that
the orbiting substructure is truncated at some tidal radius $r_{t}$
where the differential tidal force of the halo-galaxy system is equal
to the gravitational attraction of the satellite.  For non-circular
orbits the tidal radius is such that the average density of the
satellite within it is proportional to the mean density of the Galaxy
at $R_p$, the pericentric distance: $\rho_{\rm{sub}}(r_t) \propto
\rho_{\rm{gal}}(R_{p})$. N-body simulations confirm that subhalos are
truncated at the radius given by $r_t$ \citep[e.g.][]{Setal208}, but
we expect that this process will be dominant only for satellites
orbiting with large pericentric distances.


Second, the tidal field of the dark matter halo causes {\it tidal
heating} of the particles in a satellite galaxy.  This process is most
efficient for satellites on eccentric orbits \citep{GOH99}, which are
the majority according to numerical simulations, whereas it vanishes
for systems on circular orbits.  Within $\sim 20\,{\rm kpc}$ of the
Milky Way center this process destroys substructures more rapidly than
tidal stripping, owing to the strong tides that occur during close
passages of the central dark matter cusp.  We hence may also call this
process {\it halo shocking}.

Third, when an object plunges through the dense baryonic disk, it will
be subject to a process known as {\it disk shocking} \citep{OSC72,
BT87} which can, in some circumstances, produce a much stronger
response than tidal stripping \citep{TB01}.  This process is not
accounted for in dark matter only simulations, and its relevance for
the evolution of the substructure abundance has largely been ignored
thus far.  We will hence focus on this process in what follows.

Whereas a slowly varying tidal field can strip loosely bound material
through tidal stripping, a rapidly varying gravitational field, such
as that arising when a body passes through a galactic disk, will
induce gravitational shocks that add energy to an object, changing its
structure and accelerating mass loss.  For an object orbiting in the
inner regions of a galaxy on an orbit inclined with respect to the
disk, disk shocking may indeed dominate the mass loss.  This process
is known to influence the structure of globular star clusters
\citep{OSC72}, and can be more significant for dark matter
substructures.  In the solar neighborhood, the galactic disk has a
mass density $\rho \sim 0.2 \,$ M$_{\odot}{\rm pc}^{-3}$ and the
vertical scale-height with which the density decreases with distance
above or below the midplane is roughly $350\,{\rm pc}$ \citep{BT87}.
A clump of dark matter with mass of M$ \approx 10^{7}\,$ M$_{\odot}$ and
half-mass radius $r_{1/2}\approx 1\,{\rm kpc}$ has a mean density
within $r_{1/2}$ of order $\rho \sim 0.002\,$ M$_{\odot}{\rm pc}^{-3}$,
1000 times less than that of either the Milky Way disk locally or a
globular cluster with M$ \approx 10^{6}\,$ M$_{\odot}$ and
$r_{1/2}\approx 10\,{\rm pc}$.  Therefore, disk shocking can be more
important for the dynamics of dark matter substructures than for
globular clusters.

Indeed, this suggestion is supported by an additional consideration.
The characteristic internal velocities for a bound, virialized system
are $\left <v^2\right> \sim 0.4 GM / r_{1/2}$ \citep{S69, BT87},
typically $\sim 7.5\,{\rm km\, s}^{-1}$ for a globular cluster, and
only $1.4\,{\rm km\, s^{-1}}$ for a clump of dark matter of this mass.
When an object passes through a disk of vertical extent $2\,z$, its
constituents will move a distance $\Delta r \sim 2 z
\sqrt{\left<v^2\right>} / V_\perp$ along their orbits, where $V_\perp
\sim 170\,{\rm km\, s^{-1}}$ is the mean vertical speed for typical
inclined orbits through the disk of the Milky Way \citep{BT87}.  Stars
in a globular cluster will thus move a distance $\Delta r \sim$ 50 pc
during a disk crossing, implying that tightly bound stars will be
protected by adiabatic invariance \citep{BT87}, and that only the
outer, more loosely bound material will be affected.  However, because
the internal velocities characteristic of mass clumps of dark matter
are lower, the particles within these objects will move smaller
distances along their orbits, $\Delta r \sim$ 6 pc.  This is much less
than the half-mass radii of these substructures; therefore energy can
be efficiently deposited into the bulk of the mass of such an object
by disk shocking, in some cases leading to its complete disruption.

\begin{figure*}
\epsscale{1.0}
\plotone{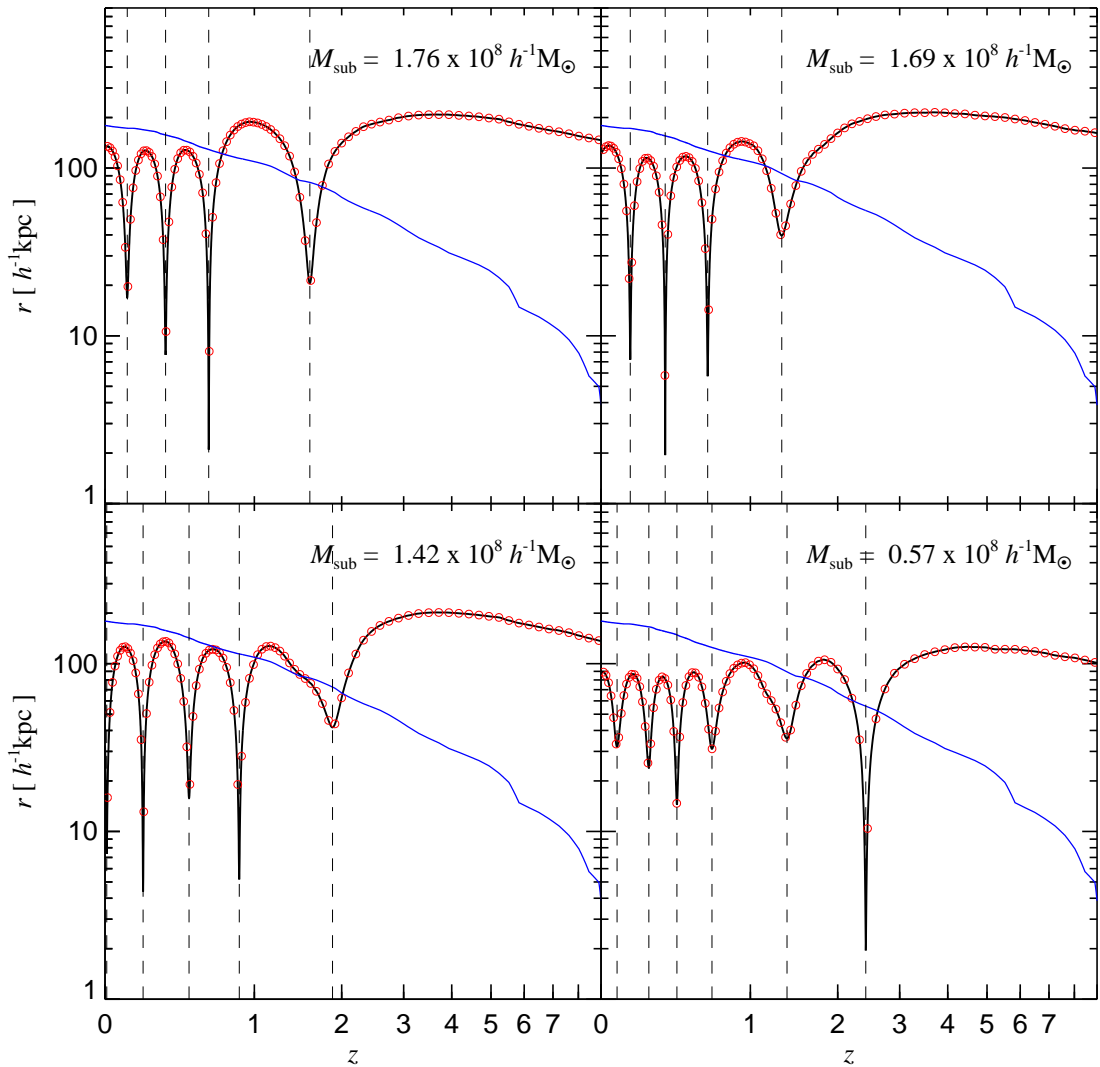}
\caption{Examples of radial orbits for four surviving substructures
  in the Aq-A-2 simulation of \citet{Setal208}. In each panel, the solid black
  line shows the physical distance of the substructure center to the center of
  the main progenitor halo of the forming Milky-Way sized halo.  The virial
  radius of this growing main halo is shown as a blue line.
  The red circles mark the output times used in the analysis; the actual
  orbit is constructed with an interpolation technique based on these output
  times.  The thin vertical lines mark the pericentric passages that are
  identified by our orbit tracking technique.
  \label{FigOrbits}}
\end{figure*}

A substructure will acquire from a disk passage an energy per unit
mass of order $\Delta E \sim 2 \left<r^2\right> g_z^2 / 3 V_\perp^2$,
where $g_z$ is the local gravitational acceleration above and below
the disk.  We note that the second moment $\left<r^2\right>$ of the
radial particle distribution of a subhalo is tightly correlated with
its half-mass radius $r_{1/2}$; from the Aquarius N-body simulations
we find $\left<r^2\right> \simeq (1.84\,r_{1/2})^2$.  The heating
energy $\Delta E$ is to be compared with the total gravitational
binding energy of the object, which we can for example estimate
according to $E_b \approx -0.28 G M^2 / r_{1/2}$, by fitting the
density profile to a \citet{H90} model, although this choice makes
little difference to our argument.  An alternative is to assume virial
equilibrium and estimate the binding energy from the total kinetic
energy in random motions of the bound subhalo, i.e.~$E_b \approx M
\left< v^2 \right> /2$.

For a substructure of mass of $10^{7}\,$M$_{\odot}$, we find $\Delta E
\sim E_b$ per orbit at 10 kpc from the center.  
If the disk mass density declines exponentially with
radius, then a corresponding estimate gives $\Delta E \approx 0.1 E_b$
per orbit at a distance of 20 kpc from the center.  
Thus, complete disruption from
disk shocking alone could occur at or within the solar radius
during a single orbit, whereas $\approx 10$ orbits would be required
to disrupt a substructure orbiting at a distance of 20 kpc.  
For objects passing
through the disk in the solar vicinity on modestly inclined or
eccentric orbits, the azimuthal period is $T_\psi \sim 2 \times 10^8$
years, implying a disruption timescale of order $\sim 2 \times 10^9$
years.  This should be regarded as an upper limit because tidal
stripping will accelerate mass loss when the substructure is above or
below the disk, and because this estimate ignores the reduction in
binding energy following each disk shock.
These calculations may overestimate the rate of depletion for  
the  subhalos with the half mass radius
larger than the disk thickness. In this case the maximum amount of energy
a substructure can receive per unit mass will be limited by the region of the
substructure that is compressed/shocked at a given time which
is of the order of disk thickness. Thus, the overall effect of disk shocking for subhalos
larger than 10${^7}$ might be overestimated in this model.

A similar estimate applied to the more massive dark matter
substructures in our simulations suggests that the disruption
timescale for these objects is also short near the disk, explaining
the rapid depletion of substructures seen in the inner regions of
Figure~\ref{FigHalos}.  Indeed, in the simulation shown in
Figure~\ref{FigHalos} with a disk, only one subhalo with maximum
circular velocity larger than $10\,{\rm km\, s^{-1}}$ survives within
$30\,{\rm kpc}$ of the center of the galaxy. As it happens, this is in
agreement with the one known Milky Way satellite on such a tightly
bound orbit, Sagittarius at $24\pm 2\,{\rm kpc}$.  Thus, it does not
appear that there is an inner missing satellite problem in the Milky
Way.

In Figure~\ref{FigEfficiency} we show the efficiency of the different
processes causing mass loss and destruction of satellites as a
function of the pericentric distances of their orbits.  For a
substructure orbiting within $20\,{\rm kpc}$ of the galactic center
the halo shocking process is stronger than tidal stripping.  However,
at slightly smaller radius, disk shocking becomes the dominant
destruction process overall.  Because of the presence of a disk in the
inner regions, the Galactic halo contracts and becomes slightly denser
in the inner regions, further enhancing the impact of tidal heating
and stripping, although we note that our estimates in
Figure~\ref{FigEfficiency} do not account for this.

\subsection{Halo shape}

Finally, we also note that the dark halo in Figure~\ref{FigHalos}
appears to be more spherical when the disk is included (in particular,
the middle and third columns of the bottom row) as compared to the
simulation without a disk (top row), in qualitative agreement with
previous studies which have tried to include baryonic material in live
halos \citep{Springel2004,Ste04a,Maccio04,Deb08,A09}.  Indeed, the halo responds to
the presence of the disk by becoming significantly more spherical.
This is quantified in Figure~\ref{FigShape} where we measure the
radial dependence of axial ratios of ellipsoidal surfaces of the halo
as a function from the halo center when a disk is included.  The
panels in Figure~\ref{FigShape} display the intermediate to major axis
($b/a$) (bottom panel) and minor to major axis ratio ($c/a$) (top
panel) of the halo measured at different distances from the center.
The presence of the disk turns the halo from a triaxial to more
spherical object within $30-40\,{\rm kpc}$ of the center, already
after $3.7\,{\rm Gyrs}$ (the disk is fully in place after 2 Gyrs).
The halo shape is not independent of radius. Indeed, this halo becomes
more round with a baryonic component only within $30-40\,{\rm kpc}$,
roughly the size of the disk and it is not affected as far out the
virial radius.  Hence, our model suggests that halos around disk
galaxies should be round only within the regions occupied by the disk
and not necessarily far beyond that.

\section{DISK PASSAGES IN COSMOLOGICAL SIMULATIONS}  \label{aquarius}

In order to extend the dynamic range of the masses of substructures
for which we can study the disk-shocking effect, we follow the orbits
of subhalos in one of the high-resolution numerical realizations of
Milky Way sized halos calculated in the `Aquarius' project
\citep{Setal08,Navarro2008}.  In this suite of dark matter only
simulations, extensive resolution studies for 6 different halos of
mass $\sim 10^{12}\,$M$_\odot$ were carried out, dark matter halos and
substructures were found at a large number of output times, and
detailed merger history trees for all these halos and subhalos were
constructed.  Using these trees allows us to accurately follow the
orbits of individual subhalos and to determine how often they are
expected to cross a fiducial disk assumed to be present in the
galaxy-sized halos.

We here focus most of our analysis on the `Aq-A-2' simulation, which has a
mass resolution of $10^4\,h^{-1}$M$_\odot$, and a virial mass of $M_{200}=
1.34\times 10^{12}\,h^{-1}$M$_\odot$ within $R_{200}=180\,h^{-1}{\rm kpc}$
(the radius enclosing a mean overdensity of 200 relative to the critical
density) for its final Milky Way-sized halo.  Based on the substructure
catalogs and merger trees, we construct continuous orbits for all
substructures by interpolating the subhalo trajectories with the discrete set
of coordinates and velocities available for the output times.  We use
reconstructed orbits based on 128 outputs between redshifts $z\sim 20$ and
$z=0$ for the analysis shown here, but we have checked that using up to 1024
outputs instead does not change the estimated disk encounter parameters in any
significant way; i.e.~the time resolution of 128 outputs is sufficient to
obtain sufficiently accurate orbital tracks for the purposes of this analysis.

In Figure~\ref{FigOrbits}, we show examples of radial orbits for four
typical substructures with mass of around $10^8\ h^{-1}$M$_\odot$ that end up
within the virial radius of the final halo.  We clearly see that the
subhalos make several orbits, sometimes reaching quite small
pericentric distances to the halo center.  Note that even
substructures that presently have a comparatively large distance from
the center may have had a close encounter with the halo center at some
earlier time.  For example, the subhalo shown in the bottom right
panel has now a distance of $r\sim 90\,h^{-1}{\rm kpc}$, but it passed
by the center at a distance of less than $2\,h^{-1}{\rm kpc}$ as early
as $z\simeq 2.5$.  Such close passages at small distances to the center
may then lead not only to significant {\em halo shocking} (which the
substructures shown in this plot in fact did survive) but also to {\em
disk shocking}.

In order to obtain an estimate for the cumulative effect of the
latter, we examine the orbits of all substructures that end up inside
the virial radius at $z=0$ and which have a mass above $M_{\rm
sub}=10^6\,h^{-1}$M$_\odot$ at the final time.  For each passage through
a fiducial plane containing the disk of the main halo (say the
$xy$-plane, but choosing a different orientation gives consistent
results), we register the distance $R$ to the center, the velocity
$V_{\perp}$ perpendicular to the disk, the subhalo's current half-mass
radius $r_{1/2}$ and its squared 3D velocity dispersion
$\left<v^2\right>$.  This allows us to characterize the strength of
the disk-shocking as
\begin{equation}
\frac{\Delta E}{E} = \frac{(1.84\,r_{1/2})^2 [4\pi G \Sigma(R)]^2}{3
  \left<v^2\right> V_{\perp}^2},
\end{equation}
where we assumed the infinite sheet approximation to relate the disk's
surface mass density $\Sigma(R)$ to the vertical gravitational field
above and below the disk.  For systems that experience multiple disk
transitions, we simply add up the different $\Delta E/E$ values.  A
significant disk shocking effect can be expected when $\Delta E/E$
reaches values of order unity.  We shall assume that $\Delta E/E\ge 1$
implies certain disruption, but we note that also smaller values,
$\Delta E/E \approxgt 0.1$, should lead to significantly accelerated
mass loss and earlier destruction.

The distribution of $\Delta E/E$ values obtained in this way is broad,
with many subhalos having very small values, suggesting that for them
disk shocking is unimportant.  However, we find that about 14.1\% of
the subhalos with mass above $M_{\rm sub}=10^6\,h^{-1}$M$_\odot$ have
$\Delta E/E$ larger than 1.0, and 24.2\% have a value larger than 0.1.

\begin{figure}
\epsscale{1.2}
\plotone{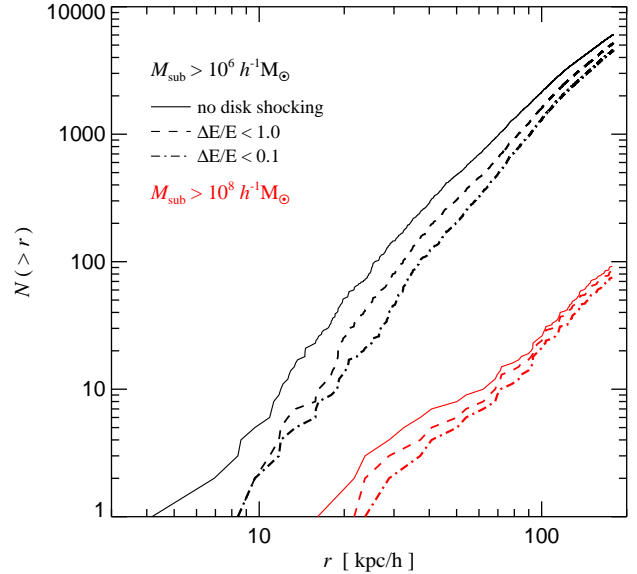}
\caption{Cumulative radial distribution of satellites with mass larger than
  $10^6 h^{-1}$M$_{\odot}$ (black lines) and larger than $10^8
  h^{-1}$M$_{\odot}$ (red lines) in the Aq-A-2 high-resolution simulation of
  the Aquarius project.  The cumulative count is shown for the original
  subhalo population (solid lines), and when subhalos are excluded for which
  the estimated disk shocking on their orbits exceeds $\Delta E/E = 1.0$
  (dashed) and $\Delta E/E = 0.1$ (dot-dashed),
  respectively. \label{FigAqCumulative}}
\end{figure}

In Figure~\ref{FigAqCumulative}, we show the estimated impact of
disk-shocking in terms of the cumulative radial count of subhalos,
comparing the counts of all subhalos to those where systems with
${\Delta E}/{E} \ge 1.0$ or ${\Delta E}/{E} \ge 0.1$ are omitted.
Interestingly, the cumulative effect is clearly stronger in the inner
than in the outer parts.  In fact, for the small mass subhalos that
dominate the $M_{\rm sub}\ge \, h^{-1}10^6\,$M$_\odot$ sample, the
reduction at radii of $\sim 30\,{\rm kpc}$ reaches a factor of $\sim
3-4$, whereas it is only about 25\% at the virial radius.  We also
include results restricted to the more massive subhalos $M_{\rm
sub}=10^8\,h^{-1}$M$_\odot$ in Figure~\ref{FigAqCumulative}.  This shows
a qualitatively similar trend, but with poorer statistics.  We find
that the more massive subhalos tend to be less affected by disk
shocking, which can be understood based on their more recent infall
time and hence smaller number of passages through the disk.

We note that the precise strength of the disk shocking effect depends,
of course, on the adopted disk model.  In the results above, we have
assumed a fixed exponential disk with scale radius $5\,h^{-1}{\rm
kpc}$ and with mass equal to 10\% of the virial mass of the final
halo.  Alternatively, we have explored models where we assumed that
the disk mass grows in proportion to the virial mass of the halo, and
the disk size scales with the halo's virial radius. This moderately
reduces the strength of the effect; we then find that 17.1\% instead
of 24.2\% of the subhalos reach a value ${\Delta E}/{E} \ge 0.1$.

Another very interesting question is whether disk shocking in
progenitor systems other than those of the main halo may be important
for the final substructure population.  Note that the results above
have only estimated disk shocking from the central disk, but perhaps
many substructures fall in as part of groups, in which case they could
already have experienced severe tidal shocks owing to the baryonic
structures in these groups.

\begin{figure}
\epsscale{1.2}
\plotone{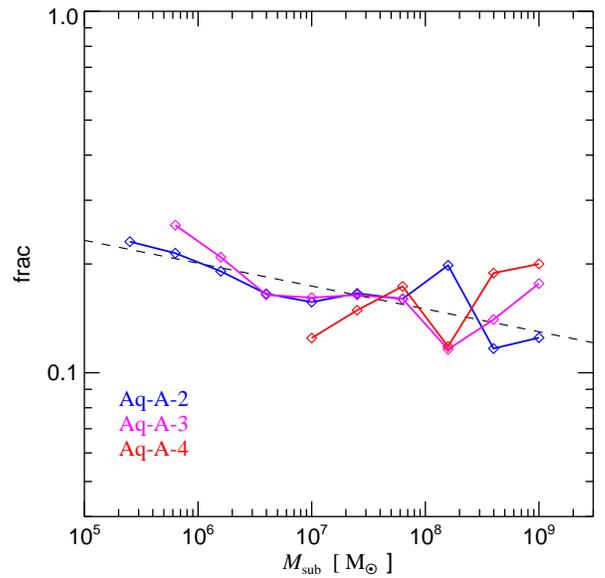}
\caption{Fraction of substructures present in the main halo at $z=0$ that were
  accreted as subhalos of another infalling structure, as opposed to being
  accreted as a main halo.  We compare results at different numerical
  resolutions.  For runs Aq-A-3 and Aq-A-4, the mass resolutions are factors of
  $3.6$ and 30 times worse than in our default simulation Aq-A-2,
  respectively.  The dashed-line marks a power-law fit to the results for
  Aq-A-2.  \label{FigAqFraction}}
\end{figure}

To investigate this idea, we first ask the question of what is the fraction of
substructures that have fallen in as a subhalo within a larger system, as
opposed to being accreted as a main halo on their own.  This can be readily
addressed with the merger trees of the Aquarius halos. In
Figure~\ref{FigAqFraction}, we show the fraction of subhalos that were
accreted as a substructure of an infalling group, as a function of their
present day mass for all subhalos in the virial radius.  We compare results
for simulations of the same halo at different numerical resolutions, finding
good convergence.  We see that the fraction of substructures that come in
already as a subhalo in a larger group is actually quite small, only about
10-20\% over the mass range considered, with a slight trend to increase
towards smaller masses \citep[see also][]{Ludlow2009}.  In hindsight this is
perhaps not too surprising, because for any given mass, there are always more
main halos as genuine subhalos, and the merger hierarchy in cold dark matter
models is actually surprisingly `shallow' \citep{Angulo2009}.

However, this also means that disk shocking in progenitor systems
different from the progenitors of the main halo (which form the `main
stem' of the merger tree) is unlikely to be very important.

We explicitly confirm this by extending our estimates of disk shocking
to include all progenitor groups and not only the main halo.  To this
end, we extract disk passage parameters also when the reconstructed
orbit of a subhalo passes by the center of a main halo different from
the main progenitor of the primary galaxy.  For simplicity, we assume
that there are also disks present in these secondary progenitor
systems, and that they always have a disk mass equal to 10\% of their
current virial mass, with a size scaled in proportion to their virial
radius.  We then find that only a relatively small number of
additional subhalos are predicted to be strongly affected by disk
shocking.  In particular, when using the ${\Delta E}/{E} \ge 0.1$
criterion, the number of affected subhalos goes up from 24.2\% to
24.8\% when the disk in the main halo is kept fixed, while it goes up
from 17.1\% to 17.8\% when the disk is assumed to always scale with
halo size.

Comparing the results from this analysis of substructure orbits in the
Aquarius halos to our direct disk-growth simulations, we find in
general reasonable agreement.  In particular, the effect estimated
from the Aquarius simulations for the inner regions of the halo is of
comparable magnitude to that found in our disk-growth experiments,
even though it appears a bit weaker overall.  But this is not
unexpected.  For one, the disk-growth simulations experience a central
compression of the dark matter halo which causes an additional
increase in the efficiency of halo shocking, an effect that we have
not taken into account in the analysis in the present section.  In
addition, disk shocking will accelerate the mass loss of surviving
subhalos, which is also an effect that we have not included here.  We
emphasize that the calculations in this section are based on
analytic estimates and involve several approximations.  A
self-consistent analysis of the orbital properties of subhalos and
their mass loss using numerical simulations when the disk is included
in a Milky Way halo is required to study these issues in more detail.

\section{DISCUSSION AND CONCLUSION} \label{discussion}

Tidal halo shocking and disk shocking are dominant effects compared to
tidal stripping for fairly eccentric orbits and have not been included
in all previous simulations of dark satellites orbiting luminous
galaxies like the Milky Way.  Our work shows that these processes can
be efficient at depleting the satellite population within $\sim$30 kpc
of the Milky Way center.

While the effects we model in our simulations can extend further than
the radial scale-lengths of disks, because many substructures are on
moderately radial orbits, our calculations do not resolve the
``missing satellite problem'' on larger spatial scales.  In principle,
the processes we have investigated here could propagate from smaller
to larger mass scales if one accounted for visible galaxies in the
substructures that merged to form the Mily Way halo, driving
pre-destruction of substructures within subhalos and leading to a
stronger depletion of satellites in the outer regions of large halos.
However, our estimates indicate that most substructures are accreted
as main halos and not as subhalos, limiting the overall importance of
this effect.  We do note, however, that if luminous satellite 
galaxies in the Milky Way formed more efficiently in 
subhaloes accreted as part of a Magellanic Group, as proposed by
\citet{DL08}, then it would not be surprising that most substructures
would remain dark, even if they are not destroyed.

Our findings have interesting implications for understanding the
nature of spiral structure in disk galaxies.  Recent simulations have
shown that dark satellites impacting a disk could play an important
role in generating the multi-armed, global spirals seen in disk
galaxies.  Our work demonstrates that the majority of the satellites
are depleted in a short time when they pass through the disk.  This
raises the question of whether an early impact of satellites inside,
or later impacts outside, the disk could still provide a source of
perturbations capable of maintaining global spiral patterns in disk
galaxies, or whether a different source is required.  Interestingly,
an alternative source might be molecular gas clumps with masses of
$10^6- 10^7\, M_{\odot}$ which are preferentially found close to the
spiral arms in the Milky Way or infalling halo clouds which have similar 
masses \citep{KH09}.

Although we do not account for the presence of live disks
we do not expect the passages of the tightly bound substructures
through the disk before these objects are destroyed to have a
significant effect on disk kinematics. \citet{Hop08} showed that the
heating done to disks by smaller perturbers is second order in the
mass ratio, so heating of the disk by substructures in the mass range
we have explored is likely to be unimportant, as also shown by 
\citet{VW99} using N-body simulations. 

Another implication of our work is that the depletion of clumps within a
few dozen of kiloparsecs from the Milky Way center might lead to a
revision of the mass fraction embedded in substructures as implied by
gravitational lensing.  These observations have been done for early-type
galaxies where there is no disk and might lead to different values when
applied to late-type/disk galaxies.  Furthermore, recent work estimating
the flux ratio anomalies in gravitational lensing owing to the presence
of substructure points out that the inner subhalo abundance predicted by
CDM-only simulations appears actually slightly too low to account for
these anomalies \citep{Xu09}.  As we have shown here, because baryonic
processes can reduce the inner substructure abundance significantly, it
might turn out that this problem is even more severe. On the other
hand, there are cases where these anomalous flux ratios are likely the
result of visible substructures. In particular, in CLASS B2045+265, one
of the most extreme anomalous flux ratio systems, \citet{McK07} find
a small dwarf galaxy companion which is sufficient to fully
explain the observed flux ratios without the need for any additional
dark substructures.

\citet{Detal08} and \citet{Setal08} have computed the dark matter
annihilation signal from the galactic substructures and from the smooth
dark matter halo of the Milky Way using very high resolution
cosmological Via Lactea II and Aquarius simulations, respectively. It
was pointed out \citep{Setal08} that the signal-to-noise for detection
of dark matter annihilation from the galactic center is in general much
larger than for any of the substructures, but in case substructures are
detectable, their typical distance from the Solar circle is predicted to
be around $20\,{\rm kpc}$. Hence this {\em detectable} substructure
population would be strongly affected by the disc shocking effect. Other
models of gamma-ray emission built on the Via Lactea II simulation
predict that the number of dark clumps detectable from the Fermi-LAT is
between 1 and 10 in five years of operation, for an optimistic dark
matter scenario \citep{Pieri09}, but see also \citet{KMS09} for an even
larger estimate assuming an additional strong boost of the annihilation
rate due to unresolved dark clumps.  Our results reduce these values by
a factor of 3 for subhalos of mass 10$^7$ M$_{\odot}$, and by a factor
of 2 for larger masses. This strengtens the expectation that detection
of the dark matter annihilation signal from the galactic center should
be easier than from nearby substructures \citep{Setal08,Pieri209}.

Ultimately disk shocking predicts that luminous satellite galaxies
orbiting through galactic disks should leave long tails of stars and
can be disrupted in a few Gyrs, a timescale much shorter than the
tails live.  This possibility is supported by recent observations of
loops of tidal tails around NGC 5907, which may be the remains of a
ghost dwarf galaxy that has not been found \citep{MD08}.

\acknowledgments

This research was partly supported by the EU Marie Curie Intra-European
Fellowship under contract MEIF-041569.  D.K. acknowledges supports by 
ITC fellowship. Numerical simulations were performed
on the Odyssey supercomputer at Harvard University. The simulations of the
Aquarius project were carried out at the Leibniz Computing Center and the
Computing Center of the Max-Planck-Society in Garching, Germany.

\bibliographystyle{apj}
\bibliography{paper}

\end{document}